\documentclass[12pt]{article}

\usepackage{fullpage}
\usepackage{hyperref}
\usepackage{graphicx}

\def\tev{\,{\rm TeV}}
\def\gev{\,{\rm GeV}}

\title{pMSSM Benchmark Models for Snowmass 2013\thanks{Work supported by the Department of Energy, Contract DE-AC02-76SF00515}}
\date{}
\author{Matthew W.~Cahill-Rowley, JoAnne L.~Hewett, Ahmed Ismail, \\ Michael E.~Peskin, and Thomas G.~Rizzo \footnote{email: mrowley, hewett, aismail, mpeskin, rizzo@slac.stanford.edu} \\\\ SLAC National Accelerator Laboratory, \\ 2575 Sand Hill Road, Menlo Park, CA 94025, USA}

\begin{document}

\begin{flushright}
\mbox{SLAC-PUB-15458}
\end{flushright}

{\let\newpage\relax\maketitle}

\begin{abstract}

We present several benchmark points in the phenomenological Minimal Supersymmetric Standard Model (pMSSM). We select these models as experimentally well-motivated examples of the MSSM which predict the observed Higgs mass and dark matter relic density while evading the current LHC searches. We also use benchmarks to generate spokes in parameter space by scaling the mass parameters in a manner which keeps the Higgs mass and relic density approximately constant.

\end{abstract}

\section{pMSSM Model Generation}

\begin{table}
\centering
\begin{tabular}{|c|c|} \hline
$m_{\tilde L(e)_{1,2,3}}$ & $100 \gev - 4 \tev$ \\ 
$m_{\tilde Q(q)_{1,2}}$ & $400 \gev - 4 \tev$ \\ 
$m_{\tilde Q(q)_{3}}$ &  $200 \gev - 4 \tev$ \\
$|M_1|$ & $50 \gev - 4 \tev$ \\
$|M_2|$ & $100 \gev - 4 \tev$ \\
$|\mu|$ & $100 \gev - 4 \tev$ \\ 
$M_3$ & $400 \gev - 4 \tev$ \\ 
$|A_{t,b,\tau}|$ & $0 \gev - 4 \tev$ \\ 
$M_A$ & $100 \gev - 4 \tev$ \\ 
$\tan \beta$ & 1 - 60 \\
$m_{3/2}$ & 1 eV$ - 1 \tev$ ($\tilde{G}$ LSP)\\
\hline
\end{tabular}
\caption{Scan ranges for the 19 (20) parameters of the pMSSM with a neutralino (gravitino) LSP. The gravitino mass is scanned with a log prior. All other parameters are scanned with linear priors.}
\label{ScanRanges}
\end{table}

Despite the continued null results from the LHC, supersymmetry in general and the MSSM in particular remain well-motivated and therefore of considerable interest to future experimental programs. We therefore introduce several benchmark points within the MSSM which predict the observed Higgs mass and dark matter relic density, yet are allowed by current experimental data. These points were taken from scans of the phenomenological MSSM (pMSSM), a subspace of the MSSM with parameters defined at the electroweak scale~\cite{CahillRowley:2012cb}. The pMSSM is defined by applying the following experimentally-motivated constraints to the R-parity conserving MSSM: ($i$) CP conservation, ($ii$) Minimal Flavor Violation at the electroweak scale, ($iii$) degenerate first and second generation sfermion masses, ($iv$) negligible Yukawa couplings and A-terms for the first two generations. In particular, no assumptions are made about physics at high scales, e.g. unification or SUSY breaking, in order to capture electroweak scale phenomenology for which a UV-complete theory may not yet exist. Imposing the constraints ($i$)-($v$) decreases the number of free parameters from 105 to 19, or 20 if the gravitino mass is included as an additional parameter. These parameters and the ranges of values considered for them in~\cite{CahillRowley:2012cb} are listed in Table~\ref{ScanRanges}.

The benchmarks presented here have been selected from two large sets of pMSSM points, henceforth referred to as ``models.'' We generated these model sets by randomly scanning the pMSSM parameter space for points with a neutralino or gravitino LSP, and checking each point for theoretical consistency as well as compatibility with precision observables, cosmology, and the 7/8 TeV LHC searches. We now describe these constraints and their application.

For each pMSSM point, we generated a spectrum via SOFTSUSY 3.1.7~\cite{Allanach:2001kg}, and compared it with SuSpect 2.41~\cite{Djouadi:2002ze}, discarding any models with theoretical inconsistencies or significant discrepancies between the two programs. We then used SUSY-HIT 1.3~\cite{Djouadi:2006bz}, modified as discussed in~\cite{CahillRowley:2012cb}, to calculate decay tables for each model. Subsequently, we employed a number of constraints from the flavor sector and precision electroweak data, specifically measurements of $(g-2)_\mu$, $b\to s\gamma$, $B_s\to\mu^+\mu^-$, $B\to\tau \nu$, $Z\to$ invisible, and $\Delta\rho$. We also imposed constraints from LEP data and model-independent limits from 2011 LHC searches for heavy stable charged particles (HSCPs) and $\phi \to \tau \tau$. For models with a neutralino LSP, we required the thermal relic density of the lightest neutralino to be below the WMAP measurement, allowing models in which the lightest neutralino does not make up all the dark matter. We additionally imposed direct detection limits from XENON 100. For models with a gravitino LSP, on the other hand, we required only that the density of gravitinos produced nonthermally through NLSP decay be below the WMAP limit. We additionally employed a host of limits arising from Big Bang Nucleosynthesis and diffuse photon/neutrino flux measurements. All of these constraints are detailed in~\cite{CahillRowley:2012cb}, and after their application, each set contains $\sim 2 \times 10^5$ surviving models. 

\begin{table}
\centering
\begin{tabular}{|l|l|l|} \hline
Search  & Energy &   Reference     \\
\hline
2-6 jets & 7 TeV & ATLAS-CONF-2012-033   \\
multijets & 7 TeV & ATLAS-CONF-2012-037    \\
1 lepton  & 7 TeV & ATLAS-CONF-2012-041    \\

2-6 jets   & 8 TeV &   ATLAS-CONF-2012-109   \\
multijets   & 8 TeV &  ATLAS-CONF-2012-103    \\
1 lepton     & 8 TeV &  ATLAS-CONF-2012-104   \\
SS dileptons & 8 TeV &  ATLAS-CONF-2012-105    \\

Gluino $\to$ stop/sbottom   & 7 TeV &   1207.4686        \\
Very light stop  & 7 TeV &    ATLAS-CONF-2012-059      \\
Medium stop  & 7 TeV &   ATLAS-CONF-2012-071  \\
Heavy stop (0l)  & 7 TeV &  1208.1447    \\
Heavy stop (1l)   & 7 TeV &  1208.2590       \\
GMSB direct stop  & 7 TeV  &   1204.6736          \\
Direct sbottom & 7 TeV  &    ATLAS-CONF-2012-106      \\
3 leptons & 7 TeV  &   ATLAS-CONF-2012-108       \\
1-2 leptons & 7 TeV  &    1208.4688          \\
Direct slepton/gaugino (2l)  & 7 TeV &   1208.2884    \\
Direct gaugino (3l) & 7 TeV  &   1208.3144      \\
HSCP      & 7 TeV  &  1205.0272    \\
Disappearing tracks  & 7 TeV  &  ATLAS-CONF-2012-111  \\
$\gamma \gamma + MET$ ($\tilde{G}$ LSP) & 7 TeV & 1209.0753 \\

\hline
\end{tabular}
\caption{Simulated LHC SUSY searches which have been applied to our pMSSM model sets. 66 (53) percent of our models with a neutralino (gravitino) LSP, including the benchmarks we present here, are not excluded by any of these searches.}
\label{SearchList}
\end{table}

Finally, we simulated the collider signatures of the remaining models to determine whether they are excluded by the current LHC data. In general, we have implemented every relevant ATLAS SUSY search available as of mid-September 2012. We generated SUSY events for each model with PYTHIA 6.4.26~\cite{Sjostrand:2006za} and PGS 4~\cite{PGS}, modified as described in~\cite{CahillRowley:2012kx} to correctly deal with gravitinos, hadronization of stable colored sparticles, and $b$-tagging. In addition, we scaled our events to NLO, calculating K-factors using Prospino 2.1~\cite{Beenakker:1996ch} for processes involving colored sparticles, and taking a constant K-factor of 1.25 for electroweak processes. Then, we implemented individual searches using a custom analysis code, following the published cuts and selection criteria as closely as possible. Our code is validated for each search, as demonstrated in~\cite{CahillRowley:2012cb}. Table~\ref{SearchList} lists the 21 SUSY searches we simulate; we also apply the most recent limits from the $\phi \to \tau \tau$ search and the observation of $B_s \to \mu^+ \mu^-$. Further results of the LHC searches may be found in~\cite{CahillRowley:2012cb, CahillRowley:2012rv, CahillRowley:2012gu}.

\section{Benchmark Points and Slopes}

\begin{figure}
\centerline{\includegraphics[height=0.4\textheight]{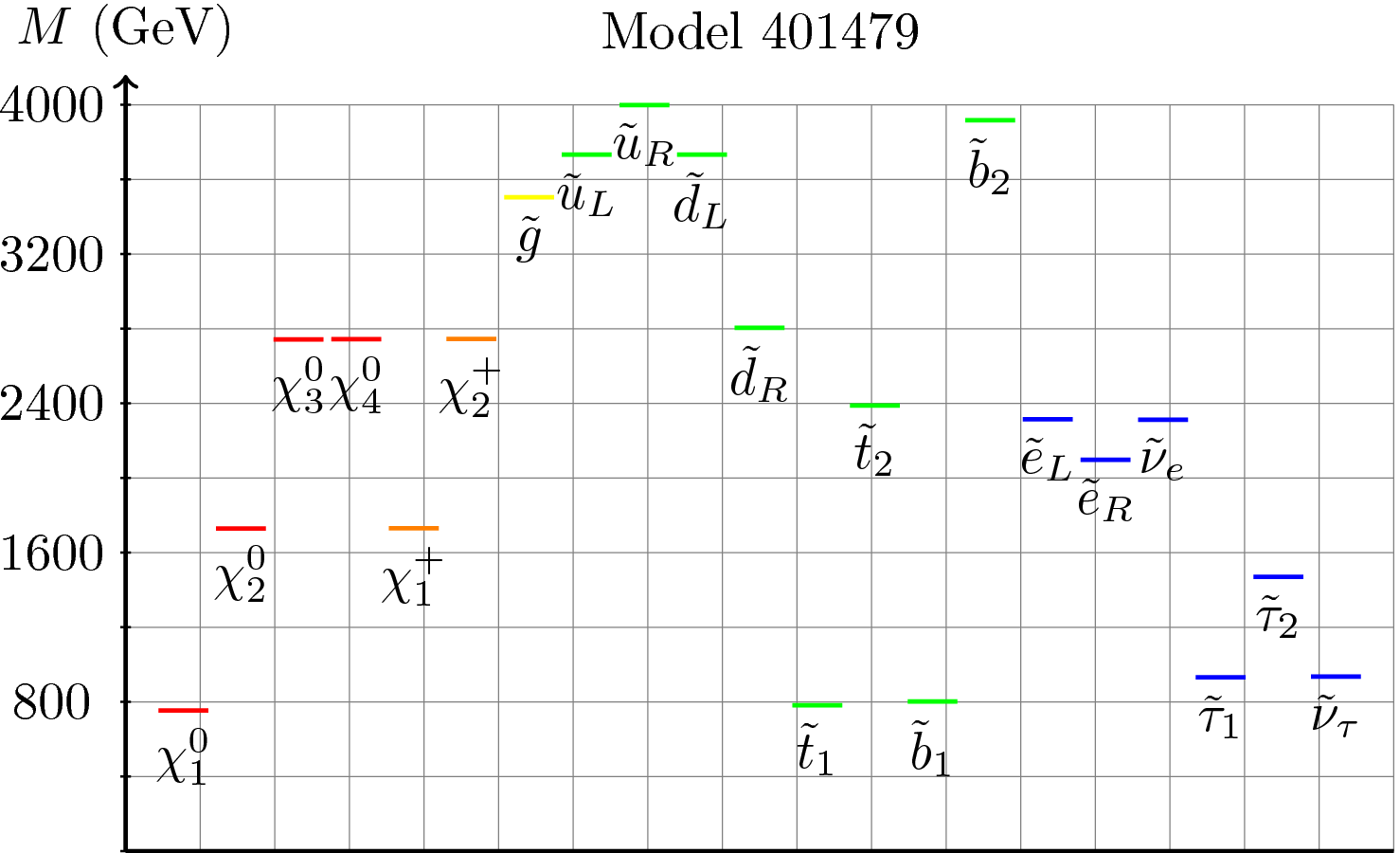}}
\caption{Bino-stop coannihilation benchmark particle spectrum.}
\label{fig:401479}
\end{figure}

\begin{figure}
\centerline{\includegraphics[height=0.4\textheight]{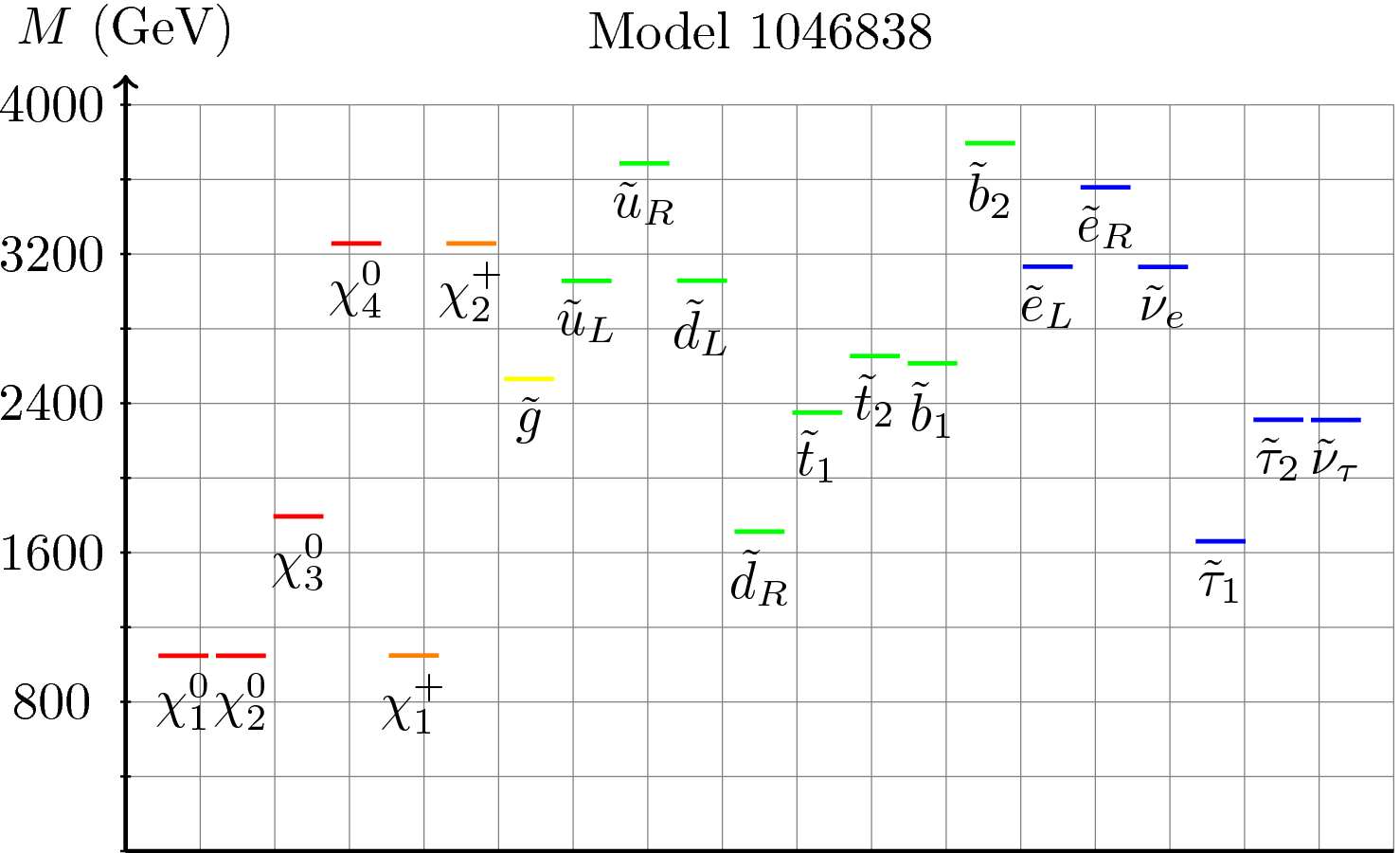}}
\caption{Pure Higgsino benchmark particle spectrum.}
\label{fig:1046838}
\end{figure}

\begin{figure}
\centerline{\includegraphics[height=0.4\textheight]{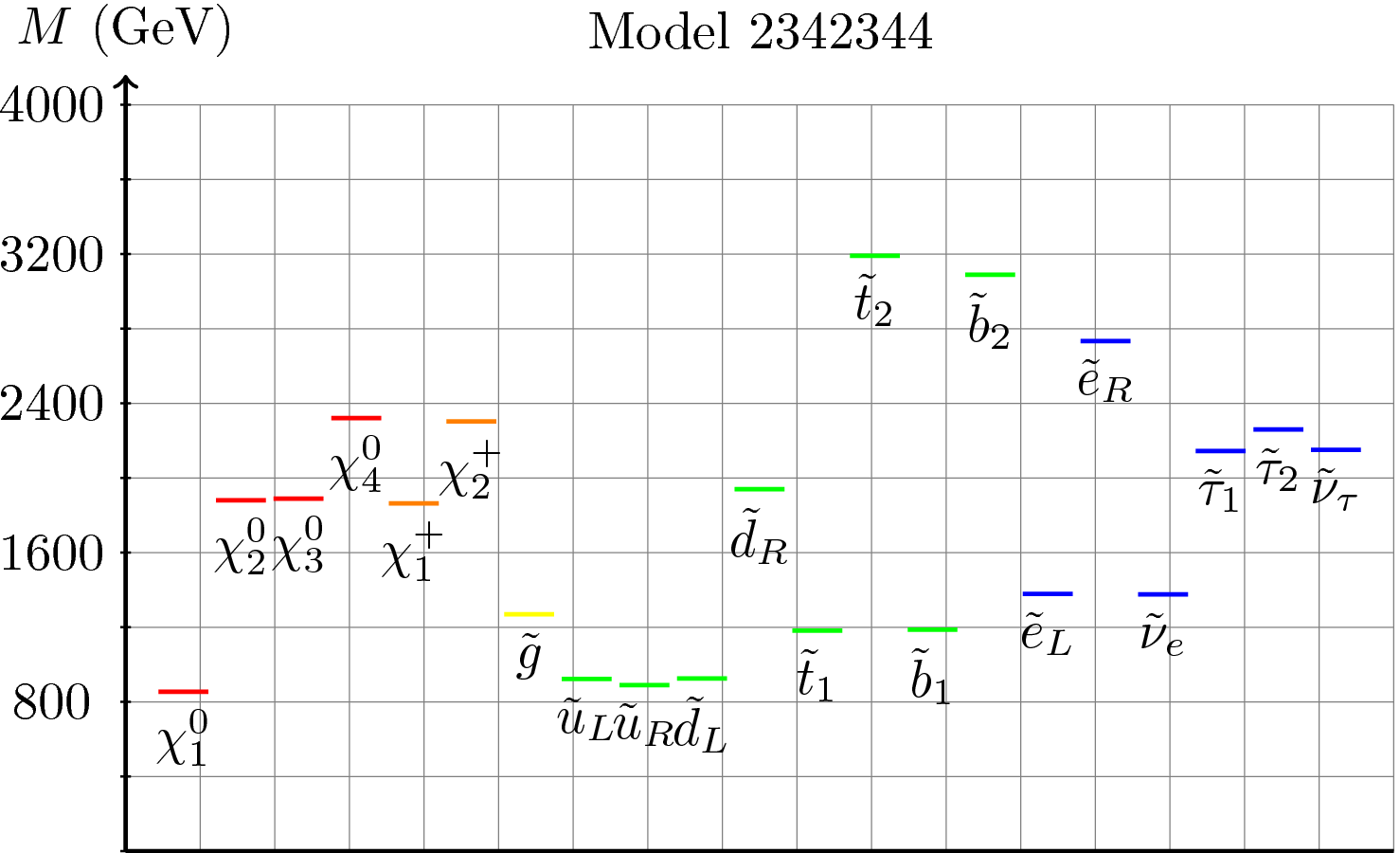}}
\caption{Bino-squark coannihilation benchmark sparticle spectrum.}
\label{fig:2342344}
\end{figure}

\begin{figure}
\centerline{\includegraphics[height=0.4\textheight]{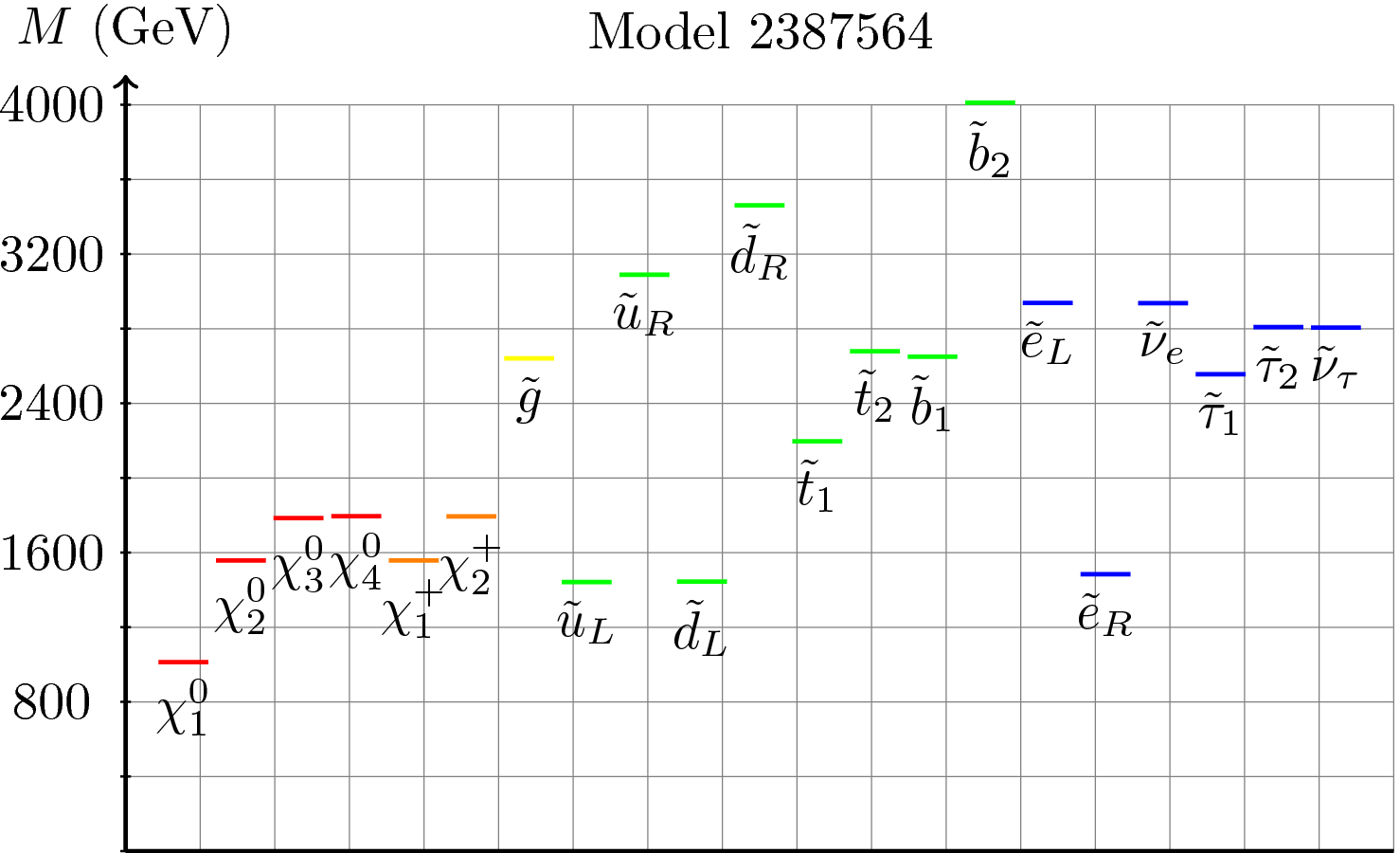}}
\caption{A funnel benchmark sparticle spectrum.}
\label{fig:2387564}
\end{figure}

\begin{figure}
\centerline{\includegraphics[height=0.4\textheight]{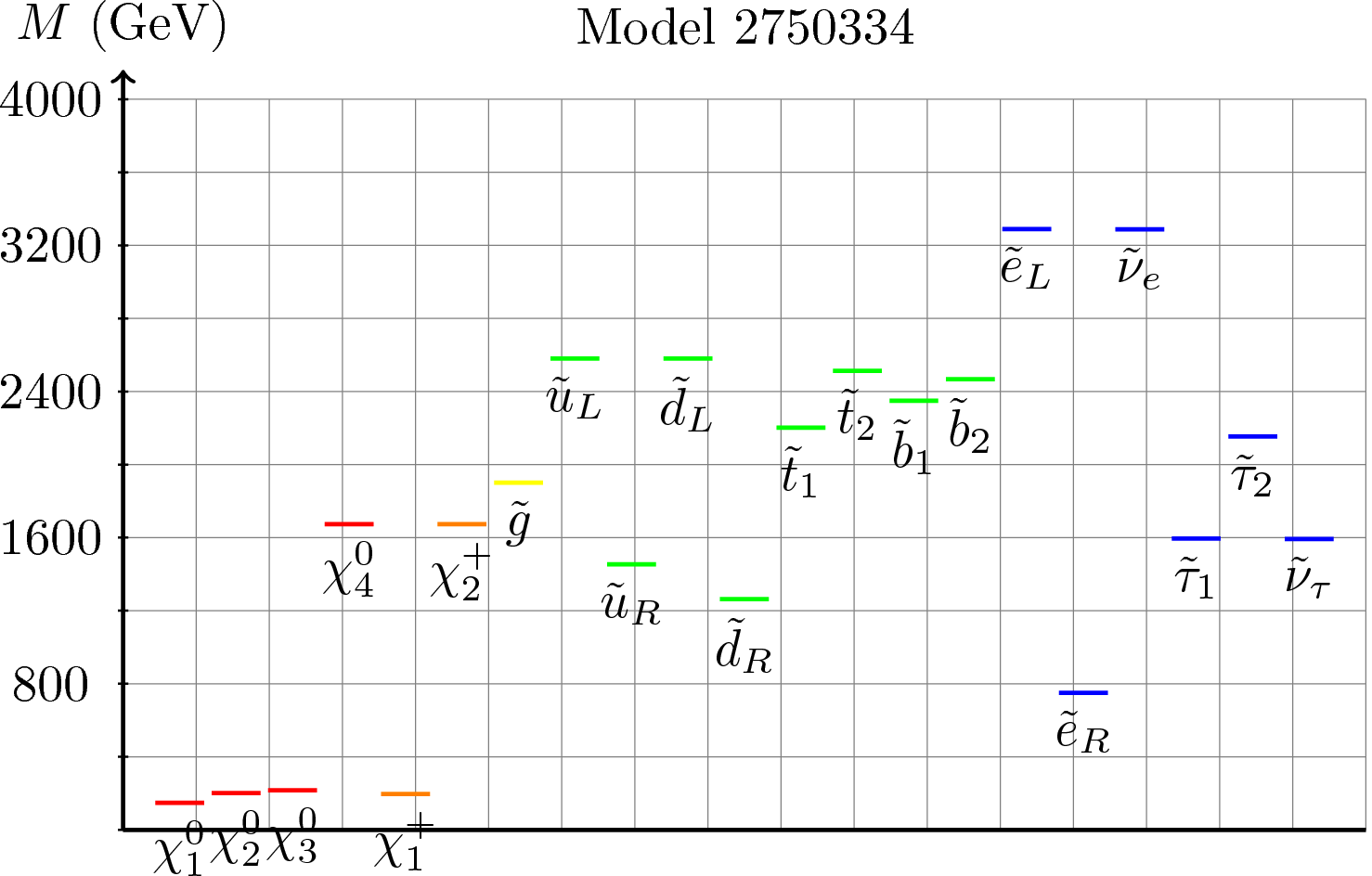}}
\caption{Well-tempered neutralino benchmark sparticle spectrum.}
\label{fig:2750334}
\end{figure}

\begin{figure}
\centerline{\includegraphics[height=0.4\textheight]{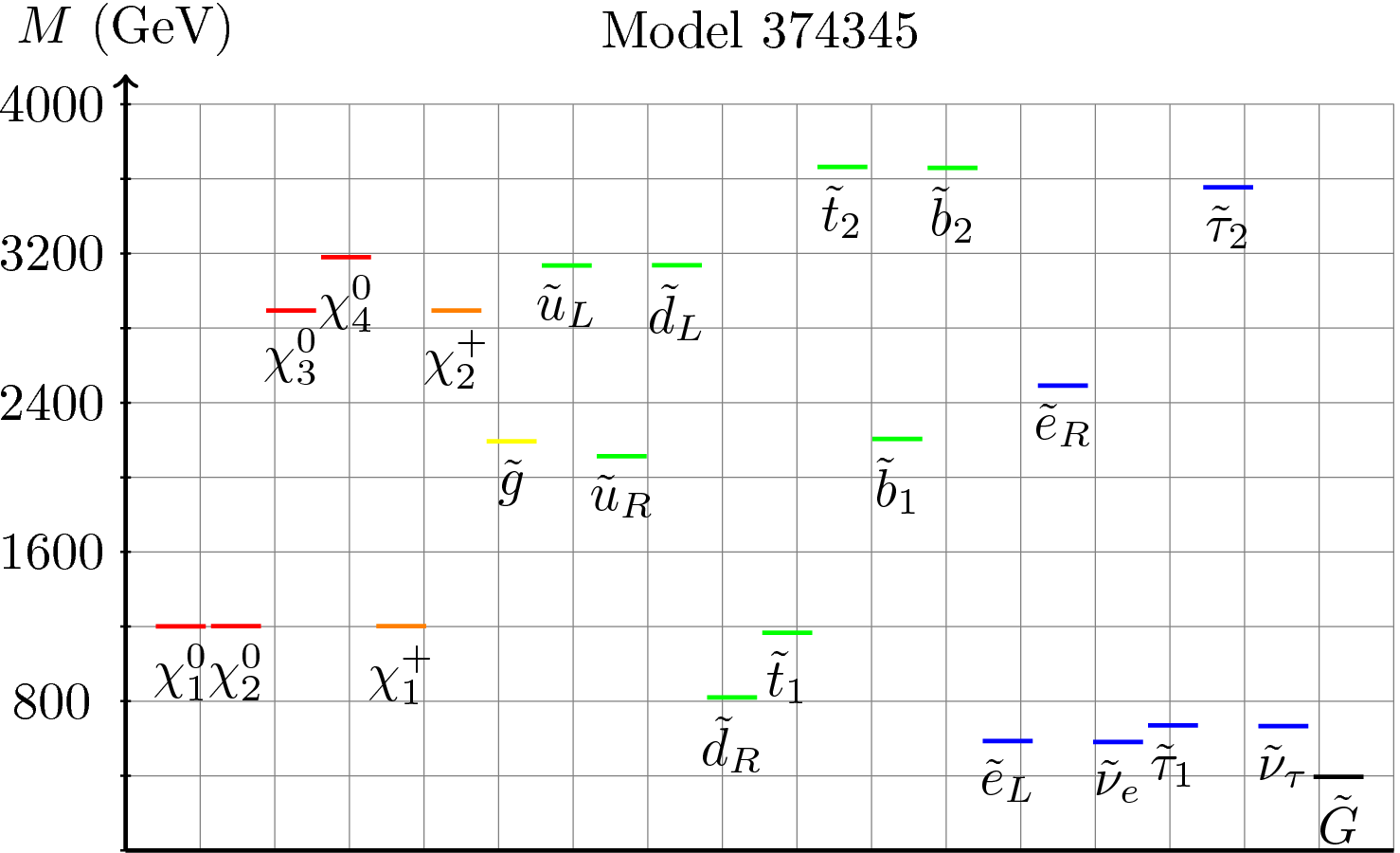}}
\caption{Gravitino super-WIMP benchmark sparticle spectrum.}
\label{fig:374345}
\end{figure}

The models that pass all constraints contain wide ranges for the light $CP$-even Higgs mass and LSP relic density. The neutralino LSP model set contains many models with $m_h$ near 126 GeV and LSP relic density close to the 2010 WMAP measurement $\Omega h^2 = 0.1126 \pm 0.0036$~\cite{Komatsu:2010fb}. Requiring that $m_h = 126 \pm 1$ GeV and that the relic density fall within $1 \sigma$ of the WMAP measurement, we are left with 24 models. These models all have significant stop mixing to give the right Higgs mass~\cite{CahillRowley:2012rv}. Notably, they contain essentially all standard mechanisms to obtain the right relic density in the MSSM, including bino-sfermion coannihilation, the well-tempered neutralino, the $A$ funnel, pure heavy Higgsinos/winos, and very heavy coannihilating Higgsinos. We propose several of these models with different dark matter scenarios as benchmarks for Snowmass 2013, with particular emphasis on their mutual relevance for the Cosmic Frontier, Energy Frontier and Intensity Frontier working groups. Brief summaries of these models and their LSPs follow. Their spectra are shown in Figures~\ref{fig:401479}--\ref{fig:374345}, and relevant information for dark matter experiments is shown in Tables~\ref{tab:dd} and~\ref{tab:id}. These may also be found at \url{http://www.slac.stanford.edu/~aismail/snowmass/index.html}.

\begin{table}
\centering
\begin{tabular}{|l|c|c|c|c|}
\hline
Model & $\sigma_\mathrm{P}^\mathrm{SI}$ (pb) & $\sigma_\mathrm{P}^\mathrm{SD}$ (pb) & $\sigma_\mathrm{N}^\mathrm{SI}$ (pb) & $\sigma_\mathrm{N}^\mathrm{SD}$ (pb) \\
\hline
401479 ($\tilde{B}$--$\tilde{t}$ coannihilation) & $1.5 \times 10^{-11}$ & $1.5 \times 10^{-9}$ & $1.5 \times 10^{-11}$ & $2.6 \times 10^{-9}$ \\
1046838 (Pure $\tilde{h}$) & $1.7 \times 10^{-10}$ & $2.5 \times 10^{-8}$ & $1.8 \times 10^{-10}$ & $1.9 \times 10^{-8}$ \\
2342344 ($\tilde{B}$--$\tilde{q}$ coannihilation) & $1.6 \times 10^{-11}$ & $4.6 \times 10^{-5}$ & $3.6 \times 10^{-11}$ & $1.1 \times 10^{-5}$ \\
2387564 ($A$ funnel) & $3.1 \times 10^{-11}$ & $1.9 \times 10^{-8}$ & $3.1 \times 10^{-11}$ & $2.6 \times 10^{-8}$ \\
2750334 (Well-tempered neutralino) & $4.3 \times 10^{-8}$ & $3.1 \times 10^{-4}$ & $4.5 \times 10^{-8}$ & $2.4 \times 10^{-4}$ \\
\hline
\end{tabular}
\caption{Direct detection cross sections for the pMSSM Snowmass benchmarks.}
\label{tab:dd}
\end{table}

\begin{table}
\centering
\noindent\makebox[\textwidth]{%
\begin{tabular}{|l|c|c|}
\hline
Model & $<\sigma v>_\mathrm{ann}$ (cm$^3$/s) & Annihilation channels \\
\hline
401479 ($\tilde{B}$--$\tilde{t}$ coannihilation) & $4.3 \times 10^{-30}$ & 79\% $\tau^+ \tau^-$, 15\% $b \bar{b}$, 5\% $g g$ \\
1046838 (Pure $\tilde{h}$) & $1.0 \times 10^{-26}$ & 50\% $W^+ W^-$, 41\% $Z Z$, 6\% $W^\pm H^\mp$, 1\% $h Z$ \\
2342344 ($\tilde{B}$--$\tilde{q}$ coannihilation) & $1.4 \times 10^{-29}$ & 69\% $g g$, 29\% $b \bar{b}$ \\
2387564 ($A$ funnel) & $3.8 \times 10^{-27}$ & 72\% $b \bar{b}$, 18\% $t \bar{t}$, 10\% $\tau^+ \tau^-$ \\
2750334 (Well-tempered $\tilde{\chi}$) & $1.9 \times 10^{-26}$ & 53\% $W^+ W^-$, 37\% $Z Z$, 5\% $h Z$, 4\% $b \bar{b}$ \\
\hline
\end{tabular}}
\caption{Annihilation cross sections for the pMSSM Snowmass benchmarks.}
\label{tab:id}
\end{table}

\begin{description}
\item[Model 401479:] Bino-stop coannihilation at $\sim 800$ GeV. The stops are difficult to see at colliders because they don't leave any MET. One might also eventually look for the stau, near 900 GeV.
\item[Model 1046838:] A pure Higgsino at 1 TeV gives the right relic density, with the lightest sfermions at 1.7 TeV.
\item[Model 2342344:] 850 GeV bino which coannihilates through many squarks at $\sim 900$ GeV. Very compressed which makes it hard to see the squarks, though the gluino is at 1.3 TeV.
\item[Model 2387564:] 1 TeV bino annihilation through the $A$. There are squarks at 1.4 TeV, reachable at colliders and with a reasonable amount of room above the LSP.
\item[Model 2750334:] Well-tempered bino-higgsino LSP at 150 GeV. The lightest sfermions are 750 GeV sleptons, with squarks significantly heavier at $\geq 1.3$ TeV. Despite the light LSP, all states below 1 TeV are uncolored.
\end{description}

The pMSSM with a gravitino LSP exemplifies the so-called ``super-WIMP'' scenario, with discovery prospects significantly better at colliders than at dark matter detection experiments. Our gravitino LSP model set contains fewer models with a heavy Higgs than the corresponding neutralino LSP model set, and the non-thermal calculation of the gravitino LSP abundance tends to only saturate the WMAP measurement when the NLSP is the sneutrino. Nevertheless, by relaxing our Higgs mass constraint above, we present one model which may serve as a super-WIMP benchmark.

\begin{description}
\item[Model 374345:] A gravitino near 400 GeV, with a detector-stable sneutrino NLSP at 580 GeV. Decays of colored sparticles should produce many final-state leptons. The light $CP$-even Higgs weighs 123.5 GeV.
\end{description}

Following previous Snowmass benchmark definitions~\cite{Allanach:2002nj}, we also identify lines in pMSSM parameter space that are potentially interesting for one-dimensional scans. In defining such ``spokes'' through our benchmarks, we require that the Higgs mass and LSP relic density are maintained as the pMSSM parameters of these benchmarks are varied. While no points along our proposed spokes have been explicitly tested against the 2012 LHC searches except for the benchmark points themselves, we generally expect that any point along such a spoke with heavier sparticles than the associated benchmark should be safe from current LHC bounds. Given the large parameter space of the pMSSM, multiple spokes may be defined for a given benchmark, allowing for the exploration of many phenomenological possibilities across the various frontiers. We note that the provided spokes are generally \emph{finite} in length, owing to the many different constraints that are applied. For example, as we construct spokes starting from the bino-squark coannihilation benchmark, the bino-squark splitting must be made smaller as the LSP mass is increased to maintain the correct relic density. Eventually, this splitting becomes small enough that the predicted direct detection cross section, mainly from $s$-channel squark exchange, grows to reach tension with current bounds from XENON 100.

After the January 2013 meeting of the BSM working group, slight modifications were made to most of these benchmark points, resulting in ``primed'' models where the 1st/2nd generation squarks and/or sleptons are close in mass. We do not anticipate that these changes will affect the viability of these benchmarks given current LHC searches. The primed models are available at \url{http://www.slac.stanford.edu/~aismail/snowmass/index.html}.

\section{Conclusion}

In this note, we have presented several benchmarks from the pMSSM that will be useful for Snowmass 2013 studies. The spectra, which we have selected from large sets of pMSSM models passing all existing experimental constraints, represent a wide variety of mechanisms for obtaining the WMAP dark matter abundance, and should give very different signatures at collider and astrophysical experiments. We have also described the construction of lines in pMSSM space that may be used for one-dimensional scans, similar to the slopes that have been defined for previous Snowmass benchmarks. Our points and slopes may be found at \url{http://www.slac.stanford.edu/~aismail/snowmass/index.html}.

\end{document}